\documentclass[prl,twocolumn,showpacs,superscriptaddress]{revtex4}
\usepackage{dcolumn}
\usepackage{amsmath}
\usepackage{amssymb}
\usepackage{amsbsy}
\usepackage{graphicx}
\usepackage{ifthen}
\newcommand{\DW}{true}

\begin{document}
\title{Fractal Substructure of a Nanopowder}

\author{Thomas Schwager}
\affiliation{Charit\'e, Augustenburger Platz 1, 13353 Berlin, Germany}
\author{Dietrich E. Wolf}
\affiliation{Universit\"at Duisburg-Essen, Fachbereich Physik, 47048 Duisburg, Germany}
\author{Thorsten P\"oschel}
\affiliation{Universit\"at Bayreuth, Physikalisches Institut, 95440 Bayreuth, Germany}
\date{\today}

\begin{abstract}
  The structural evolution of a nano-powder by repeated dispersion and
  settling can lead to characteristic fractal substructures. This is
  shown by numerical simulations of a two-dimensional model
  agglomerate of adhesive rigid particles. The agglomerate is cut into
  fragments of a characteristic size $\ell$, which then are settling
  under gravity. Repeating this procedure converges to a loosely
  packed structure, the properties of which are investigated: a) The
  final packing density is independent of the initialization, b) the
  short-range correlation function is independent of the fragment
  size, c) the structure is fractal up to the fragmentation scale
  $\ell$ with a fractal dimension close to 1.7, and d) the relaxation
  time increases linearly with $\ell$.
\end{abstract}


\maketitle

The van-der-Waals attraction between nano-particles is much stronger
than their weight. This is the reason why they agglomerate into
ramified, often fractal structures \cite{Meakin}. A well studied example are the
agglomerates formed in a filter that collects nano-particles. Once the
particle deposits have grown to micrometer size, they can be shaken
off the filter fibers easily. Collecting these rather large aerosol
flakes in a container leads to what is commonly called a nano-powder,
a fragile assembly of partly sintered micrometer flakes made of
nano-particles. Depending on the agglomeration process in the aerosol, 
as well as the influence of diffusion on the
deposition process, the nano-powder will have fractal substructures
\cite{Maedler}.
However it may be questioned, whether these are robust: 
Shaking, pouring, stirring, and all kinds of random
treatments of the container will break the nano-powder up into
fragments, presumably with a typical size determined by the prevailing
shear forces and much larger than the primary nano-particles, but not
necessarily larger than the originally collected aerosol flakes. When
allowed to settle, these fragments will reagglomerate, until the next
perturbation breaks the nano-powder up again. 

Nano-powders can be enormously porous. Porosities of more than 90 \% are
common. Since many physical properties, such as electrical conductivity,
mechanical stability, or catalytic activity, are determined by the
structure of the powder, 
it is important to know, whether there emerge robust
generic structural features as a result of repeated fragmentation and
reagglomeration processes. In this paper we present large scale
simulation results for a simple two-dimensional model which shows such
a development of a robust asymptotic structure.

\begin{figure}[t]
  \centering
 \ifthenelse{\equal{\DW}{false}}{
   \includegraphics[width=0.93\columnwidth,clip]{figs/lattice/lattice.eps}
} { \includegraphics[width=0.93\columnwidth,clip]{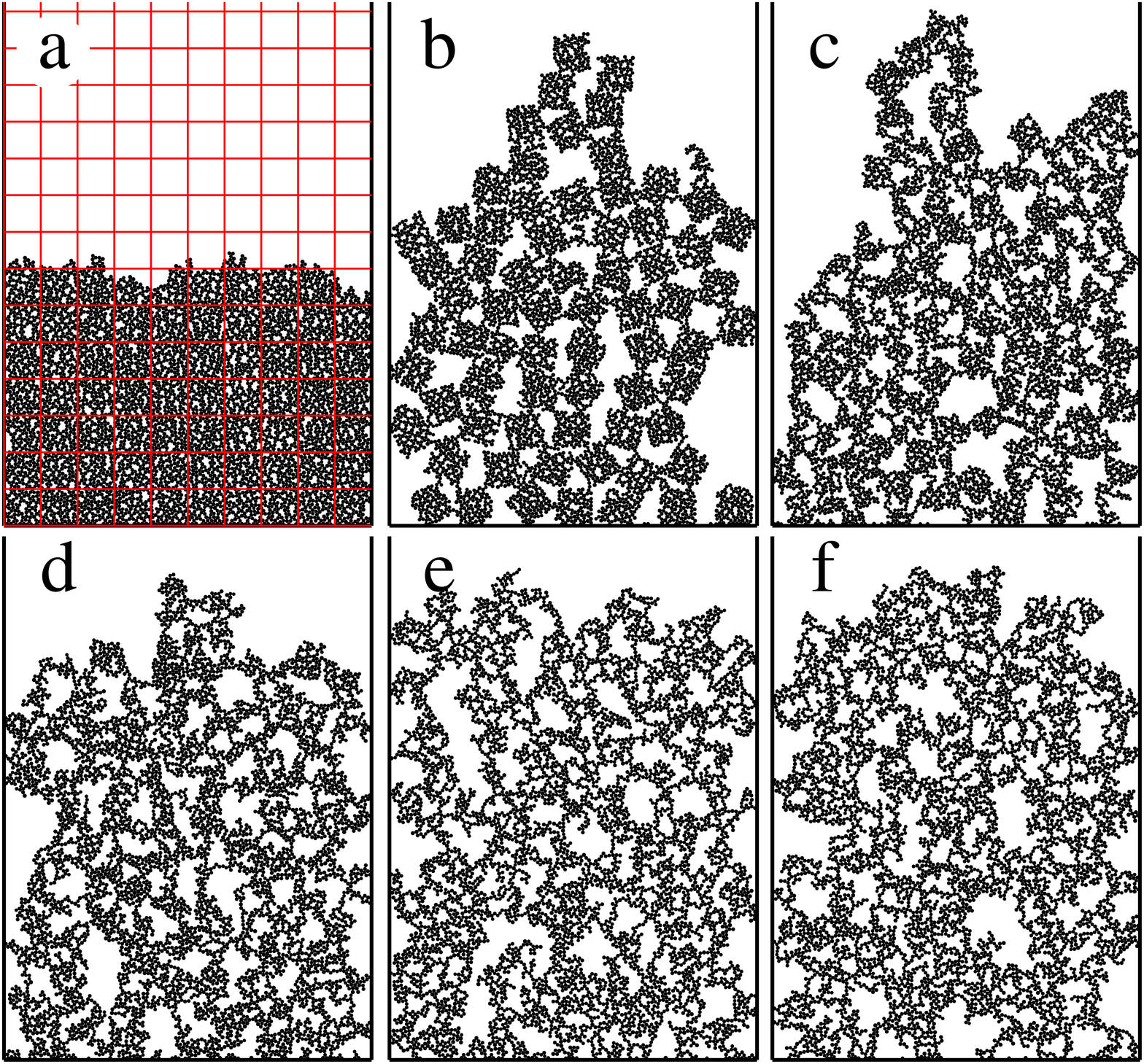}}
  \caption{Evolution of the packing of a nano-powder as described in the
    text. a) Initial packing generated by random sequential sedimentation
    \cite{VisscherBolsterli:1972}. The packing is cut by a square mesh into
    fragments ($\ell=20$); b) The fragments are considered as rigid
    bodies and deposited  
    (1$^{st}$ generation). Again the packing is cut by a square mesh (here not
    shown); c) the fragments are deposited again (2$^{nd}$
    generation), and so on; 
    d) 3$^{rd}$ generation; e) 4$^{th}$ generation; f) 120$^{th}$ generation.}
  \label{fig:generations}
\end{figure}

In our model the nano-particles are represented by up to 3 million
 discs with a narrow size distribution (10\% variance).
As initial state we take a densely packed agglomerate.  Below we
will show that the final structure is independent of the initial configuration. Then
the following procedure is repeated many times: First the agglomerate is cut
with a square mesh into portions.  The linear mesh size $\ell$ can be viewed
as the typical scale of the fragmentation process. A portion may consist of
several disconnected fragments.  These fragment flakes then settle as rigid
bodies under gravity without taking adhesion forces with other particles into
account. This is justified, if the flakes are sufficiently large, so that
their weight exceeds the van-der-Waals force between the
nano-particles \footnote{Occasionally a disconnected fragment of a
 portion can be as small as 
  a single primary particle. For such a small fragment it is not correct to
  neglect cohesion (and Brownian motion) during its settling, while assuming
  that cohesion inside larger fragments is strong enough to assure their
  stability. However, one may argue that the asymptotic structure is governed
  by the larger fragments so that the wrong dynamics for the smallest ones
  should not matter so much. Alternatively one may envisage a more subtle
  physical situation, where a freshly formed contact is only weakly cohesive,
  but cohesion becomes stronger the longer a contact exists. In this case our
  model would be justified for the smallest fragments, as well.}.
Brownian motion is neglected for the same reason.  After this
reassembly of the fragments the agglomerate is cut again with the square mesh,
and so on, see Fig.~\ref{fig:generations}.

The only model parameter is the linear mesh size $\ell$. Note that the
limit of small $\ell$ ($\ell\approx$ particle diameter) corresponds to 
non-cohesive primary particles. In any case, the cohesion forces are assumed
to be weak compared to the fragment weight, but strong enough to assure the
internal stability of the flakes.

{\em Algorithm:} To obtain statistically significant results for the structure
of nano-powders one has to consider systems with more
than one million particles for many fragmentation-reagglomeration cycles.
This is beyond the capability of Molecular Dynamics simulations.  
For our purpose we therefore generalized a model by Vis\-scher and Bolsterli
\cite{VisscherBolsterli:1972} originally intended for the sequential
deposition of macroscopic spherical particles (see
\cite{jullien93b,jullien87b,jullien88,baumann93,baumann95} for other
applications): Each particle starts at a random position well above the
already deposited material (the configuration of which is regarded as frozen
in). Following gravity, it moves downwards until it touches the bottom of the
container, where it sticks, or contacts another already deposited particle. In
the latter case it moves, again following gravity, on the surface of the
deposit until it either touches the bottom or finds a stable position in
contact with the walls and/or previously deposited particles (for more
details see \cite{algo}).  

We generalized this algorithm in order to apply it to the fragments of a
nano-powder.  In each iteration step we inspect the portions which are cut out
by the square lattice, Fig.~\ref{fig:generations}, with respect to their
connectivity. A portion may decompose into several fragments (i.e.  clusters
of connected nano-particles).  We deposit these fragments in a random sequence
in the same way as described above. A fragment rolls
down the surface of the deposit until the vertical projection of its center of
mass falls in between two points of contact.  
As in the original algorithm,
inertia is neglected, which in contrast to previous applications is less of a
problem here, because the dynamics of nano-particle flakes is usually strongly
damped.
In the following, lengths are given in units of the average particle
radius, masses in units of the particle mass, and time as number of
fragmentation-reagglomeration cycles.

{\em Asymptotic filling height.} The original Visscher-Bolsterly algorithm
produces random dense packings of spheres without fractal substructures.
Correspondingly, our generalization produces a packing of fragments that
is homogeneous on scales larger than the fragmentation length
$\ell$, as can be seen in Fig.~\ref{fig:generations}f. 
Surprisingly, however, the short range structure up to size $\ell$ develops
robust fractal properties. A first indication is given by the
$\ell$-dependence of the filling height.

\begin{figure}[t]
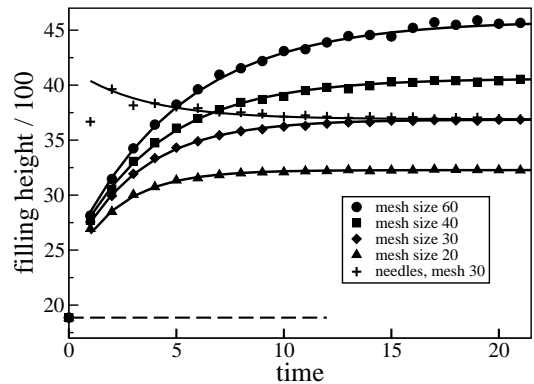

  \centering
  \ifthenelse{\equal{\DW}{false}}{
    \includegraphics[width=0.8\columnwidth]{figs/Fuellhoehe/fuellhoehe.eps}
  }{\includegraphics[width=0.8\columnwidth]{fig2.eps}}
  \caption{Evolution of filling height starting from a random
 dense packing of height $h_0$ (indicated by the dashed line). 
  The same asymptotic filling height is
  reached from above, if the particles initially form a single
  vertical needle (data marked by $+$) instead of a random dense packing. The
  full lines are fits according to Eq. \eqref{eq:asymptotic_height}. }
  \label{fig:height_top}
\end{figure}

Starting from a random dense packing
of primary particles the filling height increases towards a saturation
value.  Asymptotically, the powder adopts a very porous, statistically
invariant structure, which is robust with respect to fragmentation at
a fixed scale and subsequent gravitational settling of the fragments.
Remarkably the asymptotic filling height does not depend on the
initial configuration: Starting with all particles arranged in a
single vertical needle-like chain leads to the same value (see Fig.
\ref{fig:height_top}).  
\begin{figure}
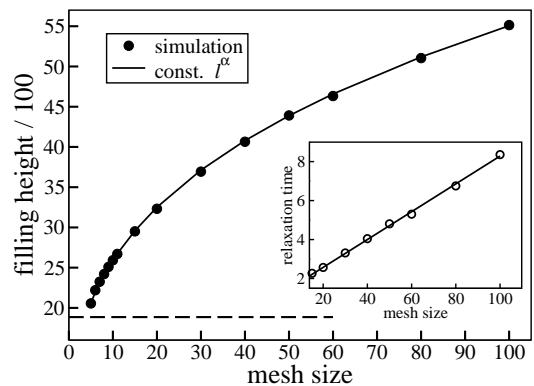

  \centering
  \ifthenelse{\equal{\DW}{false}}{
    \includegraphics[width=0.8\columnwidth]{figs/Fuellhoehe/h_of_L.eps}
  }{\includegraphics[width=0.8\columnwidth]{fig3.eps}}
  \caption{The asymptotic filling heigth $h_{\infty}$ grows as a
  power law $h_\infty(\ell)\sim\ell^\alpha$ with mesh size $\ell$. The
  full line shows the best fit, $\alpha=0.327$. Inset: The relaxation
  time $n_\text{c}(\ell)$ increases linearly with mesh size.} 
  \label{fig:height_bottom}
\end{figure}
Except for the first point
(the initial condition) the filling height $h_{n}$ at iteration step
$n$ can be fitted by an exponential approach of the asymptotic height,
$h_{\infty}$, with a relaxation time, $n_\text{c}$, 
\begin{equation}
  \label{eq:asymptotic_height}
  h_{n} = h_\infty(\ell)-\left(h_\infty(\ell)-h_0\right)
  \exp\left[-n/n_\text{c}(\ell)\right]\,. 
\end{equation}
The inset of Fig.~\ref{fig:height_bottom} shows that
\begin{equation}
n_\text{c}(\ell) \propto \ell^z \quad \text{with} \quad z=1.
\end{equation}
For the asymptotic filling height, a power law 
\begin{equation}
h_{\infty}(\ell) \propto \ell^{\alpha} \quad \text{with} \quad \alpha = 0.327
\end{equation}
gives a very good fit(see Fig.~\ref{fig:height_bottom}).

\begin{figure}[t]
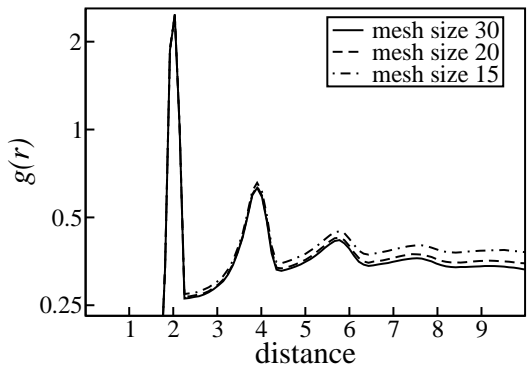

  \centering
  \ifthenelse{\equal{\DW}{false}}{
    \includegraphics[width=0.8\columnwidth,angle=0]{figs/Autocorrelation/autocorr_log.eps}
}{\includegraphics[width=0.8\columnwidth,angle=0]{fig4.eps}}
  \caption{Pair correlation function (semi-log-plot) of the steady state
   structure.}
  \label{fig:auto}
\end{figure}
This implies that the number of portions cut from the steady state
configuration of a system of width $L$ scales like $N_\text{p} = h_{\infty}L/\ell^2 \propto
\ell^{\alpha-2}$. Consequently, the mass per portion is 
$M/N_\text{p} \propto \ell^{d_\text{f}}$ with
\begin{equation} 
d_\text{f} = 2-\alpha = 1.67 \pm 0.03.
\label{eq:d_f}
\end{equation}
One can interpret $d_\text{f}$ as the fractal dimension of the structure on
length scales smaller than the fragmentation length $\ell$, because the short
range part of the pair correlation function $g(r)$ 
does not depend on $\ell$ (see
Fig.~\ref{fig:auto}). This means that the short range structure 
is independent of the overall density. For large distances, on the
other hand, the pair correlation function approaches the overall density,
which decreases with increasing $\ell$.

\begin{figure}[t]
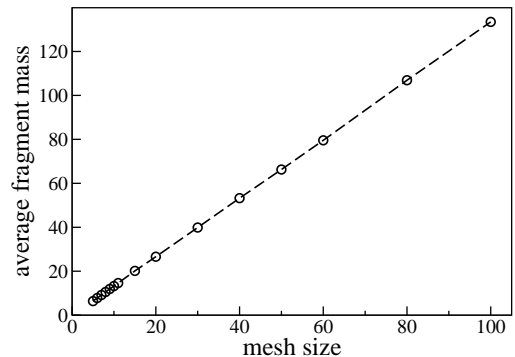

   \centering
   \ifthenelse{\equal{\DW}{false}}{
     \includegraphics[width=0.8\columnwidth,bb=15 34 719 531]{figs/Clustersize/cluster-final-insetA.eps}
}{\includegraphics[width=0.8\columnwidth,bb=15 34 719 531]{fig5.eps}}
  \caption{Average fragment mass
    as a function of the mesh size $\ell$ in the asymptotic steady
    state.} 
  \label{fig:mean_fragment_size_over_mesh}
\end{figure}

On first glance, however, the fractal dimension Eq.~\eqref{eq:d_f}
seems at odds with the fact, that the asymptotic average fragment mass
grows linearly with the mesh size $\ell$
(Fig.~\ref{fig:mean_fragment_size_over_mesh}),  suggesting,
instead, that the fragments are effectively one-dimensional
structures.  This puzzle can only be resolved by assuming that the number
of disconnected fragments $N_\text{f}$ per mesh cell has itself a power
law dependence on $\ell$:  
\begin{equation}
\frac{N_\text{f}}{N_\text{p}} \propto \ell^{\beta}.
\label{eq:def_beta}
\end{equation}

As we are going to prove in a moment, the
fragmentation-reagglomeration dynamics implies that the exponents
$\beta$ and $d_\text{f}$ must be related by
\begin{equation}
\beta = d_\text{f} -1.
\label{eq:beta_d_f}
\end{equation}
Consequently the mass per fragment is linear in $\ell$:
\begin{equation}
\frac{M}{N_\text{f}} \propto \frac{\ell^{d_\text{f}}}{\ell^{\beta}}
\propto \ell.
\label{eq:mass_per_fragment}
\end{equation}

The proof of the scaling relation Eq.~\eqref{eq:beta_d_f} is based on
the steady state condition that
the deposition of fragments reestablishes on average as
many contacts as were cut in the preceding fragmentation step.  An
$\ell\times\ell$- mesh cell contains $\propto \ell^{d_\text{f}}$ particles, of
which $\propto \ell^{d_\text{f}-1}$ are at the cell boundary. Hence, the number
of particle contacts cut by the boundary of one cell scales as
\begin{equation}
  N_\text{cut} \propto \ell^{d_\text{f}-1}\ .
\end{equation}
Each of the $\ell^{\beta}$ fragments forms two new contacts when
deposited. Equating $N_\text{cut}=2 N_\text{f}/N_\text{p}$ gives
Eq.~\eqref{eq:beta_d_f}.

\begin{figure}[t]
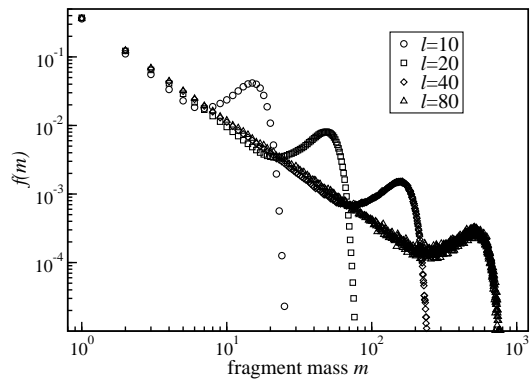

  \centering
  \ifthenelse{\equal{\DW}{false}}{
    \includegraphics[width=0.8\columnwidth,angle=0,clip]{Groessenverteilung/FIGS/allallfrags.eps} 
}{\includegraphics[width=0.8\columnwidth,angle=0,clip]{fig6.eps}}
  \caption{Normalized fragment mass distribution for different
  mesh sizes $\ell$. $f(m)$ is the number of fragments of mass $m$
  divided by the total number of fragments for a given $\ell$.}   
  \label{fig:cluster_sizes}
\end{figure}
{\em Fragment mass distribution.} A detailled understanding of the
fragment properties is provided by the distribution of fragment masses, shown
in Fig. \ref{fig:cluster_sizes}. It reveals that that one must distinguish two
types of fragments, large chunks at the upper end of the mass spectrum with a
characteristic size $m_\text{c}$, and scale invariant dust responsible for the
power law part that is cut off by $m_\text{c}$. Comparing the mass
distributions for different mesh sizes $\ell$ shows, that they can
approximately be written in the form
\begin{equation}
  \label{eq:cluster_size}
  f(m,\ell) = m^{-\tau} \tilde{f}\left(\frac{m}{m_\text{c}(\ell)}\right)\ ,
\end{equation}
where the scaling function $\tilde{f}(x)$ is constant for $x\ll 1$, goes through a
maximum at $x=1$, and has an approximately Gaussian tail for $x\gg 1$. 
\begin{figure}[t]
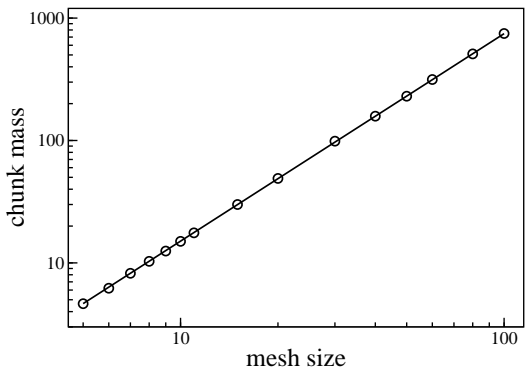

  \centering
  \ifthenelse{\equal{\DW}{false}}{
    \includegraphics[width=0.8\columnwidth,clip]{Groessenverteilung/FIGS/fragment_size.eps}
}{\includegraphics[width=0.8\columnwidth,clip]{fig7.eps}}
  \caption{Chunk mass $m_c$ as a function of mesh size. Slope of straight line
  is 1.695.}
  \label{fig:chunk_mass}
\end{figure}
The typical mass $m_\text{c}$ of the chunks has a power-law dependence on the
mesh size, $m_\text{c} = 0.304~\ell^{1.695}$ (Fig.~\ref{fig:chunk_mass}), the
exponent being in good agreement with the value of $d_\text{f}$,
Eq.~\eqref{eq:d_f}.

\begin{figure}[t]
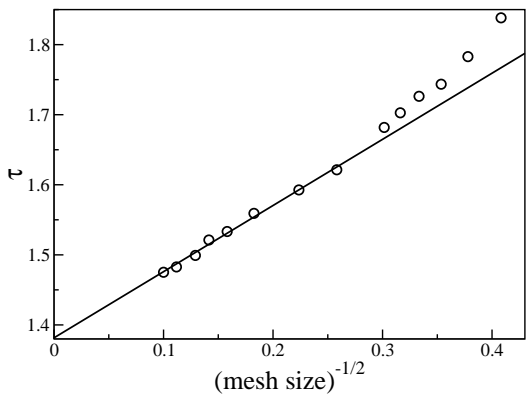

  \centering
  \ifthenelse{\equal{\DW}{false}}{
    \includegraphics[width=0.8\columnwidth,clip]{Groessenverteilung/FIGS/tauOver1bySQRTl.eps}
}{\includegraphics[width=0.8\columnwidth,clip]{fig8.eps}}
  \caption{Effective dust exponent $\tau$ vs. $1/\ell$.}
  \label{fig:exponent_tau}
\end{figure}
The evaluation of the dust exponent, $\tau$, is more difficult, since the
slope fitted to the power law part of the mass distribution in the
log-log-plot Fig.~\ref{fig:cluster_sizes} decreases significantly with
increasing $\ell$. An extrapolation of $\tau$ for $\ell^{-1/2} \rightarrow
0$, see Fig.~\ref{fig:exponent_tau}, gives an estimate of $\tau \approx 1.38$.
However, an independent method of determining $\tau$ gives a larger value. It
is based on the important observation that the width of the
chunk-distribution is proportional to $m_\text{c}$. Hence the fraction of
chunks among the fragments vanishes like 
\begin{equation}
f_\text{chunks} \propto m_\text{c}^{1-\tau} \propto \ell^{d_\text{f}(1-\tau)},
\label{eq:f_chunks}
\end{equation}
because $\tau$ is larger than 1. Hence, the normalization of
$f(m)$ for $\ell\rightarrow\infty$ implies that
\begin{equation}
1=
f(1)\sum_{m} m^{-\tau}=
f(1)\ \zeta(\tau) 
\end{equation}
with Riemann's zeta-function at the argument $\tau$.
Solving this equation numerically with $f(1)\approx 0.36$ (for the
$\ell$-values we considered) gives an estimate $\tau \approx 1.46$.

We have seen, that the overwhelming number of fragments are dust particles,
apart from a vanishing fraction of chunks. However, these dust particles carry
only a vanishing fraction of the total mass $M$. According to
Eq.~\eqref{eq:mass_per_fragment}, 
\begin{equation}
m_\text{dust} \leq f(1)\ \zeta(\tau-1)\frac{N_\text{f}}{M} \propto
\frac{1}{\ell} 
\end{equation}
vanishes for $\ell\rightarrow\infty$. Essentially all the mass is in the few
chunks. This explains, why the mass (essentially mass of chunks) per fragment
(essentially per dust particle) has nothing to do with the fractal dimension.

Now a consistent picture has formed: Each portion (or mesh cell) contains
typically one chunk. The number of fragments per portion, which according to
Eq.~\eqref{eq:def_beta} scales like $\ell^{\beta}$, can thus be
identified with $1/f_\text{chunks}$, which according to Eq.~\eqref{eq:f_chunks}
scales as $\ell^{d_\text{f}(\tau-1)}$. This shows that the fractal dimension
of the chunks and the dust exponent are not independent of each other. Using
Eq.~\eqref{eq:beta_d_f} they obey the scaling relation
\begin{equation}
d_\text{f}(2-\tau)=1\ .
\label{eq:scaling_relation}
\end{equation}
For $d_\text{f}=1.695$ this implies $\tau=1.41$, in between the two
$\tau$-values obtained above. 

We would like to thank I. Goldhirsch for fruitful discussions. This
work was supported by the German-Israeli Foundation by grant
no. I-795-166.10/2003.


\end{document}